\documentclass[12pt,a4paper]{article}
\usepackage{amsmath}
\usepackage{amsfonts}
\usepackage{amssymb}
\usepackage{cite}
\usepackage[pdftex]{graphicx}

\title{\Large{Possible physical self-asserting of the  homogeneous  vector potential}\\
\large{A testing puzzle based on  a G.P. Thomson-like arrangement}}
\author{Spiridon  Dumitru\\ 
\normalsize{(Retired)Department of Physics, \textit{"Transilvania"} University,}\\
\normalsize{B-dul Eroilor 29, 500036 Brasov, Romania} \\ \normalsize{E-mail: s.dumitru42@unitbv.ro}}

\date{\today}

\begin{document}

\maketitle

\begin{abstract}

\small{It is suggested a testing puzzle  able to reveal the self-asserting property of the homogeneous vector potential field. As pieces of the puzzle are taken three reliable entities : (i) influence of a potential vector on the  de Broglie wavelength (ii) a G.P. Thomson-like experimental arrangement and (iii) a  special coil designed  to create a homogeneous vector potential. The alluded property is  not connected  with  magnetic fluxes  surrounded by the vector potential field lines, but it  depends on  the fluxes which are outside of the respective lines. Also the same  property shows that in the tested case the vector potential  field is  uniquely defined physical quantity, free of  any adjusting gauge. So the  phenomenology of the suggested quantum test  differs on that of macroscopic theory where the vector potential is not uniquely defined and allows a gauge adjustment. Of course that the proposed test has to be subjected to adequate experimental validation.}

\end{abstract}
PACS Codes: 03.75.-b ,03.65.Vf, 03.65.Ta, 03.65.Ca, 06.30 Ka

Keywords: Homogeneous vector potential, de Broglie wavelength,  G.P. Thomson-like experiment, Electrons diffraction fringes, Outside magnetic fluxes, Needlessness of an adjusting  gauge.

\newpage
\section{Introduction}

The physical self-asserting (objectification) of the vector potential $\vec A$ field, distinctly of electric and/or magnetic  local actions, is known as Aharonov-Bohm Effect (ABE). It aroused scientific discussions for more than half a century ( see \cite{1,2,3,4,5,6,7,8} and references).  As a rule in  ABE context the vector potential    is  curl-free field, but it is non-homogeneous (\textbf{n-h}) i.e. spatially non-uniform. In the same context the alluded self-asserting is connected quantitatively with    magnetic fluxes  surrounded by the lines of $\vec A$ field.  In the present paper we try to suggest a testing puzzle intended to reveal the possible physical self-asserting property of a homogeneous (\textbf{h}) $\vec A$ field,. Note that in both \textbf{n-h} and \textbf{h} cases here we consider only fields which are constant in time.

The announced puzzle has as constitutive pieces   three reliable  Entities(\textbf{E})  namely :  
    
 $\bullet$  $\textbf{E}_1$: The fact that a potential vector change the values  de Broglie wavelength $\lambda^{dB}$ of  electrons. $\blacksquare$ 
    
  $\bullet$   $\textbf{E}_2$: An experimental arrangement of G. P. Thomson type,  able  to monitor   the mentioned $\lambda^{dB}$ values.$\blacksquare$
     
 $\bullet$   $\textbf{E}_3$: A feasible special coil designed so as to create a \textbf{h}-$\vec A$ field. $\blacksquare$\\
 
 Accordingly,  in its wholeness,  the puzzle has  to put together the mentioned entities and, consequently, to synthesize a clear verdict regarding the alluded  property of a \textbf{h}-$\vec A$ field .
  
     Experimental setup of the  suggested   puzzle is  detailed in the next Section 2. Esential theoretical considerations concerning  the action of   a \textbf{h}-$\vec A$ field  are given in Section 3. The above noted considerations are fortified in Section 4 through a set of numerical estimations for the quantities aimed to be measured by means of the puzzle.  Some concluding thoughts regarding a possible positive result of the suggested puzzle  close the principal body of the paper in Section 5. Constructive and computational details regarding the special coil designed to generate a \textbf{h}-$\vec A$ field  are presented in the 	Appendix.
\section{Setup details  of the experimental puzzle}
The setup of the suggested experimental puzzle is pictured and detailed below in Fig.~\ref{fig:1}.  It consists in a G. P. Thomson-like arrangement partially located in an area with a \textbf{h}-$\vec A$ field . The alluded arrangement is inspired from some illustrative figures \cite{9,10} about  G. P. Thomson's original experiment and it disposes in a straight line the following elements: electrons source, electrons beam ,  crystalline grating and detecting screen.  An area with a \textbf{h}-$\vec A$ field can be obtained through a certain special coil whose constructive and computational details are given in the alluded Appendix.
\begin{figure}[h]
\centering
\includegraphics[width=0.9\columnwidth]{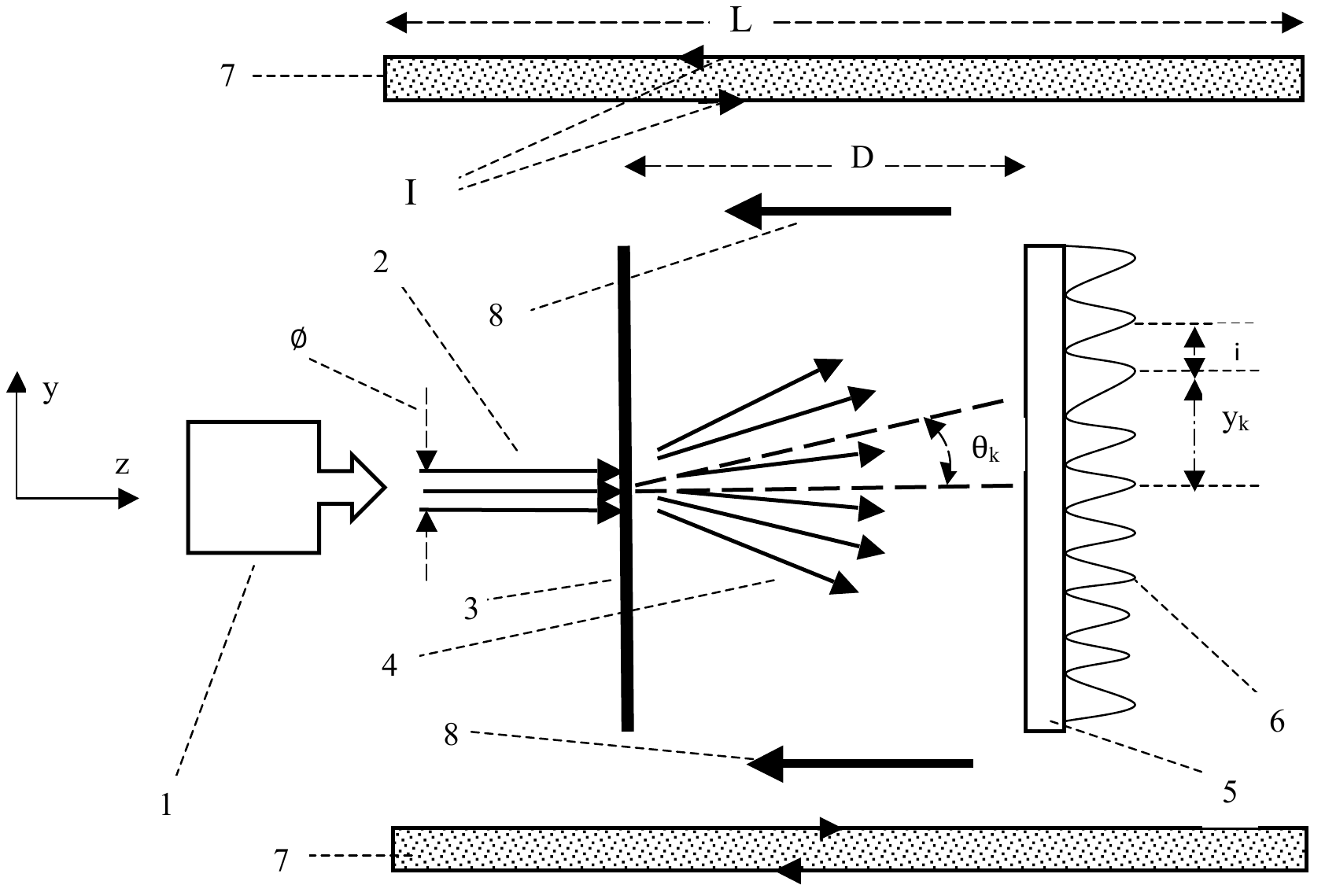}
\caption{\textbf{Plane section in the image of suggested experimental \newline puzzle setup, accompanied by the following explanatory records} \newline
1 -- Source for a beam of  mono-energetic and parallel moving  electrons; 2 --Beam of electrons in parallel movements; 3 –- Thin crystalline foil as diffraction grating; 4 -- Diffracted electrons; 5 -- Detecting screen; 6 –Fringes in the plane section of diffraction pattern ;  7 -- Special coil able to create a $\mathbf h-\vec A$ field;  8 -- $\mathbf h-\vec A$ field ; $\phi$ = the width of the electrons beam with $\phi \gg a$  ($a$ = interatomic spacing in crystal lattice of the foil -3); $\theta_k$ = diffraction angle for the $k$-th order fringe ($k=0,1,2,3,\ldots$) ; $y_k$ = displacement from the center line of the $k$-th order fringe ; $i$ = interfringe width = $y_{k+1} - y_k$; $D$ = distance between crystalline foil and screen ($D\gg \phi$); $L$ = length of  the special coil ($L\gg D$) ; $I$ = intensity of current in wires of the coil. 
}
\label{fig:1}
\end{figure}

The explanatory records  accompanying Fig.~\ref{fig:1} have to be supplemented with the next notes:

 $\bullet$ \textbf{Note 1} : If in Fig.~\ref{fig:1}  are omitted  the elements 7 and  8   ( i.e. the sections in special coil  and  the lines of \textbf{h}-$\vec A$   field ) one  obtains  a  G. P. Thomson-like arrangement as it is illustrated   in references \cite{9,10}.$\blacksquare$
 
 $\bullet$ \textbf{Note 2 }: Evidently the above mentioned G. P. Thomson-like arrangement is so designed and constructed that it can be placed inside of a vacuum glass container. The respective container is not showed in Fig.~\ref{fig:1} and it  will  leave out the special coil.$\blacksquare$
 
 $\bullet$ \textbf{Note 3 } : At the incidence on  crystalline foil  the electrons beam must ensure a coherent and plane front of de Broglie waves. Similar ensuring is required \cite{11} for optical diffracting waves at the incidence on a  classical diffraction grating. $\blacksquare$
 
 $\bullet$ \textbf{Note 4 }: In Fig.~\ref{fig:1} the detail 6 displays only  the linear projections of the fringes  from the diffraction pattern. In its wholeness the respective pattern consists in a set of concentric circular fringes  (diffraction rings). $\blacksquare$

\section{	Theoretical considerations concerning \\action of a \textbf{h}-$\vec A$  field}

The leading idea of the above suggested puzzle is to search the possible  changes caused by a \textbf{h}-$\vec A$  field in diffraction of  quantum (de Broglie) electronic  waves. That is why now  firstly  we  remind some  quantitative characteristics of the diffraction phenomenon. 

The most known scientific domain  where the respective phenomenon  is studied regards the optical light waves \cite{11}. In the respective domain one uses as main element   the so called \textit{"diffraction grating"} i.e. a piece   with  periodic structure having  slits separated by  distances $a$ and  which  diffracts the light into  beams in different directions. For a light  normally incident on such an element the grating equation (condition for intensity maximum )  has the form : $a \cdot \sin \theta _k  = k\lambda$, where $ k=0,1,2,...$. In the respective equation   $\lambda$ denotes  the light wavelength and $\theta_k$  is the angle at which the diffracted light  have  the $k-th$ order maximum. If the diffraction pattern is received on a detecting screen the $k-th$ order maximum appear on the screen in position $y_k$ given by relation $\tan \theta _k  = (y_k/D)$, where $D$ denote the  distance between screen and grating. For  the distant screen assumption, when $D>>y_k$,  can be written  the relations: $sin \theta _k  \approx \tan \theta _k \approx(y_k /D) $. Then, with regard to the mentioned assumption,  one obtains that diffraction pattern  on the screen is characterized by an  interfringe distance  $i= y_{k+1} - y_k$ given through the relation
\begin{equation}\label{eq:1}
i = \lambda \frac{D}{a}
\end{equation}

Note the fact that the above quantitative aspects of diffraction have a generic character, i.e. they are valid for all kinds of waves including the de Broglie ones. The respective fact is presumed  as a  main element of the experimental puzzle suggested in the previous section.  Another main element of the alluded puzzle is the largely agreed  idea \cite{1,2,3,4,5,6,7,8} that the de Broglie electronic  wavelength $\lambda^{dB}$  is influenced by the presence of a $\vec A$ field. Based on the two before mentioned main elements the considered puzzle can be detailed as follows.
 
 In experimental setup depicted in Fig.~\ref{fig:1} the  crystalline  foil 3 having interatomic spacing $a $ plays the role of a diffraction grating.  In the same experiment on the detecting screen 5 is expected to appear a diffraction pattern of the electrons. The respective pattern would be characterized by an  interfringe distance  $i^{dB}$ definable through the formula $i^{dB}= \lambda^{dB}\cdot(D/a)$. In that formula   $D$ denote distance between crystalline foil  and screen, supposed to satisfy the condition  $ D >> \phi $),
  where $\phi$ represents the the width of the incident electrons beam. In absence of a \textbf{h}-$\vec A$ field the $\lambda^{dB}$ of a non-relativistic electron is known as having the expressions:
\begin{equation}\label{eq:2}
\lambda ^{dB}  = \frac{h}{p_{mec}} = \frac{h}{{mv}} = \frac{h}{{\sqrt {2m{\rm \mathcal{E}}} }}
\end{equation}
In the above expressions $h$ is the Planck's constant while $p_{mec}$, $m$, $v$ and $\mathcal{E}$ denote respectively the mechanical momentum, mass, velocity and kinetic energy of the electron. If the alluded energy is obtained in the  source of electrons beam (i.e. piece 1 in Fig.~\ref{fig:1}) under the influence of an  accelerating voltage $U$ one can write $\mathcal{E}=e\cdot U$ and $p_{mec} = mv = \sqrt {2meU} $. 

Now, in connection with the situation depicted in Fig.~\ref{fig:1}, let us look for  the expressions of the electrons characteristic  $\lambda^{dB}$ and respectively of $i^{dB}= \lambda^{dB}\cdot(D/a)$ in presence of a \textbf{h}-$\vec A$  field. Firstly we note the known fact \cite{6} that a particle with the electrical charge $q$ and the mechanical momentum $\vec p_{mec} = m\vec v$ in a potential vector $\vec A$  field acquires an additional ($add$) momentum, $\vec p_{add} = q \vec A$, so that its \textit{"effective" (eff)} momentum is $\vec P_{eff} = 
\vec p_{mec} + \vec p_{add} = m\vec v + q\vec A$. Then for the electrons ( with $q = - e$) supposed to be implied in the experiment depicted in Fig.~\ref{fig:1} one obtains the effective (eff) quantities
\begin{equation}\label{eq:3}
\lambda _{eff}^{dB} \left( A \right) = \frac{h}{{mv + eA}}\;;\quad i_{eff}^{dB} \left( A \right) = \frac{{hD}}{{a\left( {mv + eA} \right)}}
\end{equation}
Further on we have to take into account the fact that the \textbf{h}-$\vec A$  field acting in the discussed experiment is generated by a special coil whose plane section is depicted by the elements  7 from Fig.~\ref{fig:1}. Then from the relation \eqref{eq:10} established in Appendix we have $A =\mathcal{K}\cdot I $, where $\mathcal{K} = \frac{{\mu _0 N}}{{2\pi }} \cdot \ln \left( {\frac{{R_2 }}{{R_1 }}} \right)$.  Add here the fact that in the considered experiment $mv= \sqrt {2meU} $. Then for the effective interfringe distance  $i_{eff}^{dB} $  of diffracted electrons one find
\begin{equation}\label{eq:4}
i_{eff}^{dB} \left( A \right) = i_{eff}^{dB} \left( {U,I} \right) = \frac{{hD}}{{a\left( {\sqrt {2meU}  + e\mathcal{K}I} \right)}}
\end{equation}
respectively
\begin{equation}\label{eq:5} 
\frac{1}{{i_{eff}^{dB} \left( {U,I} \right)}} = f\left( {U,I} \right) = \frac{{a\sqrt {2me} }}{{hD}}\sqrt U  + \frac{{ae\mathcal{K}}}{{hD}}I
\end{equation}

 \section{  A set of numerical estimations}
 
The verisimilitude of the above suggested testing  puzzle can be  fortified to some extent by transposing several of the previous formulas in their corresponding numerical values. For such a transposing firstly we will appeal to numerical values known from G.P. Thomson-like experiments. So, as regards the elements from Fig.~\ref{fig:1} we quote the values $a = 2.55\cdot10^{-10} m$ ( for a crystalline foil of copper) and $D=0.1 m$. As regards $U$ we   take the often quoted value:  $U =30 \cdot kV$. Then the mechanical momentum of the electrons  will be $p_{mec} =mv= \sqrt {2meU}=9.351\cdot10^{-23} kg\cdot m\cdot s^{-1} $. The additional (add) momentum of the electron, induced by the special coil, is of the form $p_{add}=e\mathcal K \cdot I$ where $\mathcal{K} = \frac{{\mu _0 N}}{{2\pi }} \cdot \ln \left( {\frac{{R_2 }}{{R_1 }}} \right)$. In order to  estimate the value of $\mathcal K$ we propose the following practically workable values: $R_1= 0.1 m$, $R_2= 0.12 m$, $N = 2 \pi R_1 \cdot n$ with $n = 2\cdot 10^3 m^{-1}$ = number of wires (of $1 mm$ in diameter) per unit length, arranged in two layers. With the well known values for $e$ and $\mu_0$ one obtains
$p_{add} =7.331\cdot 10^{-24}(kg\cdot m\cdot s^{-1}\cdot  A^{-1})\cdot I$ (with $ A = ampere$). 

For  wires of $1 mm$ in diameter, by changing the polarity of voltage powering the coil,  the current $I$ can be adjusted in the range $I \in (-10\ to \ +10 ) A $. Then the effective momentum  $\vec P_{eff} = \vec p_{mec} + \vec p_{add}$ of the electrons have the values within the interval 
$(2.040\ to\  16.662)\cdot10^{-23} kg\cdot m \cdot s^{-1}$. Consequently, due to the above mentioned values of $a$ and $D$, the  effective interfringe distance  $i_{eff}^{dB} $ defined in \eqref{eq:4} changes in the range $ (1.558\; to\; 12.725) mm$, respectively its inverse from \eqref{eq:5} has values within the interval $(78.58 \;to \;641.84) m^{-1} $.

Now note that in  absence of \textbf{h}-$\vec A $  field (i.e. when $I=0$) the interfange distance $i^{dB} $ specific to a simple G.P. Thomson experiment has the value $i^{dB}  = \frac{{hD}}{{a\sqrt {2meU} }} = 2.776 mm$. Such a value is within the values range  of $i_{eff}^{dB} $ characterizing the presence of a \textbf{h}-$\vec A $ field. This means that the quantitative evaluation  of the mutual relationship of $i_{eff}^{dB} $ versus $I$ and therefore of the self-asserting of a  \textbf{h}-$\vec A $ field can be done with  techniques and accuracies similar to those  for simple G.P. Thomson experiment.

\section{Some concluding remarks}

The  aim  of the experimental puzzle  suggested above is to test a possible physical self-asserting for a \textbf{h}-$\vec A $  field. Such a test can be done concretely by  comparative measurements of the interfringe  distance $i_{eff}^{dB}$ and of the current $I$. Additionally it must to examine  whether the results of the mentioned measurements verify the relations \eqref{eq:4} and \eqref{eq:5} ( particularly according to \eqref{eq:5} the quantity $(i_{eff}^{dB})^{-1}$ is expected to show a linear dependence of $I$). If the above outcomes are positive one can be notified the fact that  a \textbf{h}-$\vec A $  field has its own characteristic of physical self-asserting.   Such a fact leads in one way or another to the following remarks (\textbf{R}): 

 $\bullet$   $\textbf{R}_1$: The self-asserting of \textbf{h}-$\vec A $  field differs from the  one of  \textbf{n-h}- $\vec A $  field which appears in ABE. This because, by comparison with the illustrations from \cite{12} , one can see that : \textit{(i)} by changing of \textbf{n-h}- $\vec A $ the diffraction pattern undergoes a simple translation on the screen,  without any modification of interfringe distance, while \textit{(ii)} according to the relations \eqref{eq:4} and \eqref{eq:5} a change of \textbf{h}-$\vec A $ (by means of current $I$) does not translate the diffraction pattern but varies the associated interfringe distance. The mentioned variation is similar with those induced \cite{12} by changing (through accelerating voltage $U$) the values  of mechanical momentum $\vec p_{mec} = m\vec v$ for electrons. $\blacksquare$
 
 $\bullet$   $\textbf{R}_2$ : There are a difference between the objectification (self-asserting)  of \textbf{h}-$\vec A $  and \textbf{n-h}-$\vec A $ fields in relation with the   magnetic fluxes surrounded or not  by the  field lines.  The difference is pointed out by the next aspects: 
 
 \textit{(i)} On the one hand, as it is known from ABE,  in case of a \textbf{n-h}-$\vec A $  field  the self-asserting is in a direct dependence on magnetic fluxes surrounded   by the  field lines.
 
 \textit{(ii)} On the other hand the self-asserting of a \textbf{h}-$\vec A $  field is  not connected  with  magnetic fluxes surrounded by the  field lines. But note that due to the relations \eqref{eq:4} and \eqref{eq:5} the respective  self-asserting  appears to be  dependent (through the current $I$)  on   magnetic fluxes not surrounded by field lines of the \textbf{h}-$\vec A $.  $\blacksquare$
 
  $\bullet$   $\textbf{R}_3$ : Another particular characteristic  of the  self-asserting  forecasting above for \textbf{h}-$\vec A $ is that  in the proposed test the vector potential field  appears as an uniquely defined physical quantity free of  any adjusting gauge. So the  phenomenology of the suggested test  differs on that of macroscopic situations where \cite {13,14} the vector potential is not uniquely defined and allows a gauge adjustment.  Surely that such a fact (difference) and its implications have to be approached in more elaborated studies.

\subsubsection*{\small {Postscript}} 

As presented above the suggested puzzle and its positive result appear as purely hypothetical things, despite of the fact that they are based on the  essentially reliable entities (constitutive pieces) presented in Introduction. Of course that a true confirmation of the alluded result  can be done by an action of  putting in practice the whole puzzle. Unfortunately I do not have access to material logistics able to allow me an effective practical test of the puzzle in question. That is why I warmly appeal  to  experimentalist researchers that have adequate logistics to put in practice the suggested test and to verify its validity. 

\section*{	\Large{Appendix}\\
\large{Constructive and computational details  for a special coil 
able to create a \textbf{h}-$\vec A $  field}} 

\subsubsection*{ \normalsize{The case of an ideal coil}}

An experimental area of macroscopic size with a \textbf{h}-$\vec A $  field can be realized with the aid of  a special coil whose constructive and computational details are presented below.  The announced details are improvements of ideas promoted by us in an early preprint \cite{15}. 

The basic element in designing of the mentioned coil is the \textbf{h}-$\vec A $  field generated by a rectilinear infinite conductor  carrying a direct current.  If the conductor is located along the axis Oz and current have the intensity I, the Cartesian components ( written in SI units) of the mentioned \textbf{h}-$\vec A $  field are given \cite{16} by formulas:
\begin{equation}\label{eq:6}
A_x \left( 1 \right) = 0\quad \quad A_y \left( 1 \right) = 0\quad \quad A_z \left( 1 \right) =  - \mu _0 \frac{I}{{2\pi }}\ln r
\end{equation}
Here $r$ denote the distance from the conductor of the point where \textbf{h}-$\vec A $ is evaluated and $\mu _0 $ is  vacuum permeability.

Note that formulas  \eqref{eq:6} are of ideal essence because they describe a \textbf{h}-$\vec A $  field generated by an infinite (ideal) rectilinear conductor. Further firstly   we will use the respective formulas in order to obtain  the  \textbf{h}-$\vec A $  field generated by an ideal annular coil. Later one we will specify the conditions in which the results obtained for  the ideal coil can be used with good approximation in the characterization of  a real  ( non-ideal)  coil of practical interest for the puzzle-experiment suggested and detailed in Sections 2,3 and 4.
\begin{figure}[h]
\includegraphics[width=0.9\columnwidth]{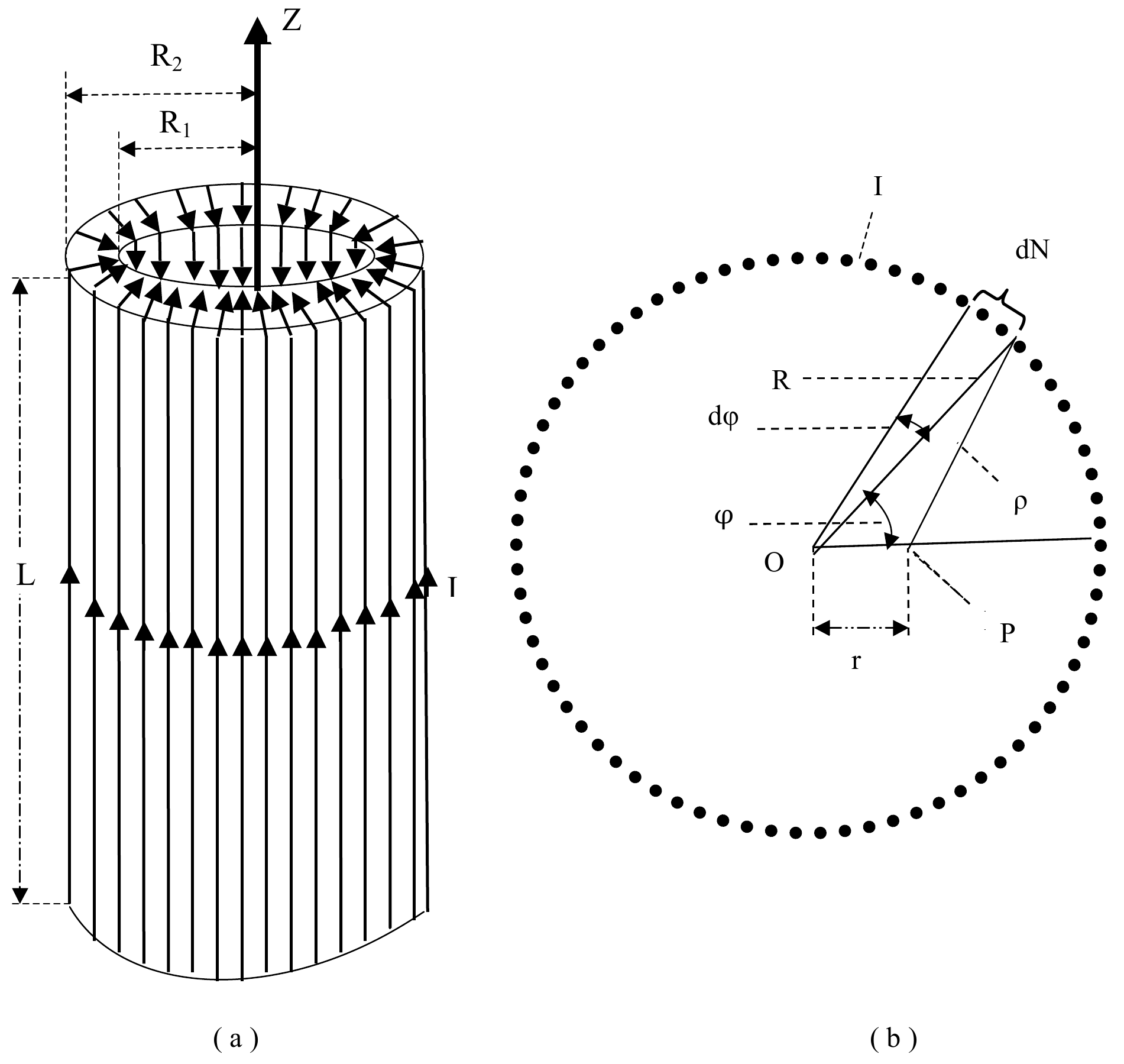}
\caption{Schemes for an  annular  special coil}
\label{fig:2}
\end{figure}

The  mentioned special coil has the shape depicted in Fig.~\ref{fig:2}-(a) (i.e. it is a toroidal coil of rectangular section). In the respective figure the finite quantities  $R_1$ and $R_2$  represent the inside  and outside finite radii of coil while $L\rightarrow\infty$ is the length  of the coil. For evaluation of the \textbf{h}-$\vec A $ generated  inside of the mentioned coil let us now consider an  array of infinite rectilinear  conductors carrying direct currents of the same intensity I. The conductors are mutually parallel and uniformly disposed on the circular cylindrical surface with the radius $R$. Also the conductors are parallel  with $Oz$ as symmetry axis. In a cross section the considered array are disposed on a circle of radius $R$ as can be seen in Fig.~\ref{fig:2}-(b). On the respective circle the azimuthal angle $\varphi$ locate the infinitesimal element of arc whose length is $Rd\varphi$. On the respective arc is placed a set of   conductors whose number is $dN = \left( {\frac{N}{{2\pi }}} \right)d\varphi $, where $N$ represents  the total number of conductors in the whole considered array. Let be an observation  point $P$ situated at distances $r$ and $\rho$ from the center $O$ of the circle respectively from the infinitesimal arc (see the Fig.~\ref{fig:2}-(b) ). Then, by taking into account \eqref{eq:6}, the z-component of the \textbf{h}-$\vec A $  field generated in $P$ by the dN conductors is given by relation
\begin{equation}\label{eq:7}
A_z \left( {dN} \right) = A_z \left( 1 \right)dN =  - \mu _0 \frac{{NI}}{{4\pi ^2 }}\ln \rho  \cdot d\varphi 
\end{equation}
 where  $\rho  = \sqrt {\left( {R^2  + r^2  - 2Rr\cos \varphi } \right)} $ . Then the all $N$ conductors will generate in the point $P$ a \textbf{h}-$\vec A $  field whose value $A$ is
\begin{equation}\label{eq:8}
A =  A_z \left( N \right) =  - \mu _0 \frac{{NI}}{{8\pi ^2 }}\int\limits_0^{2\pi } {\ln \left( {R^2  + r^2  - 2Rr\cos \varphi } \right)}  \cdot d\varphi 
\end{equation}
For calculating the above integral can be used formula (4.224-14) from \cite{17}. So one obtains
\begin{equation}\label{eq:9}
A =  - \mu _0 \frac{{NI}}{{2\pi }}\ln R
\end{equation}
This relation shows that the value of $A$ does not depend on $r$, that is on the position of $P$ inside the circle of radius $R$. Accordingly this means that inside the respective circle the potential vector is homogeneous. Then starting from \eqref{eq:9},  one obtains that the inside space of an ideal annular coil depicted in Fig.~\ref{fig:2}-(a) is characterized by a \textbf{h}-$\vec A $  field whose value is 
\begin{equation}\label{eq:10}
A = \mu _0 \frac{{NI}}{{2\pi }}\ln \left( {\frac{{R_2 }}{{R_1 }}} \right)
\end{equation}

\subsubsection*{\normalsize{From the ideal coil to a real one} }

The above presented coil is of ideal essence because   their  characteristics were evaluated on the base of ideal formulas \eqref{eq:6}. But in practical matters, such is the puzzle-experiment proposed in Sections 2 and 3,  one needs of  a real coil which may be effectively constructed in a laboratory. That is why it is important to specify the main conditions in which  the above ideal results   can be used in real situations. The mentioned conditions are displayed here below.

$\bullet$ \textit{On the geometrical sizes } : In  laboratory it is not possible to operate with objects of infinite sizes. Then it must to note the  restrictive conditions so that  the characteristics of the ideal  coil discussed above to remain as good approximations for a real coil of similar geometric form. 
In the case of a finite coil having the form depicted in the Fig.~\ref{fig:2}-(a) the alluded restrictive conditions impose the relations $L>>R_1$, $L>>R_2$ and $L>>(R_2 - R_1)$. If the respective coil is regarded as a piece in the puzzle-experiment from Fig.~\ref{fig:1} there  are indispensable the relations $ L  >> D $ and $L >> \phi $.

$\bullet$ \textit{ About the marginal fragments} :  In principle  the marginal fragments of coil  (of widths  $(R_2 - R_1)$) can have disturbing effects on the Cartesian components of  $\vec A$ inside the the space of practical interest. Note that, on the one hand, in the above mentioned  conditions $L>>R_1$, $L>>R_2$ and $L>>(R_2 - R_1)$ the alluded effects can be neglected in general practical affairs. On the other hand in the particular case of the  proposed coil the alluded effects are also diminished by the symmetrical flowings  of currents in the respective marginal fragments.
  
$\bullet$ \textit{As concerns the helicity} :  The discussed annular coil is supposed to be realized by turning a single piece of wire. The spirals of the respective wire are not strictly parallel with the symmetry axis of the coil ( $Oz$ axis) but they have a certain helicity (corkscrew-like path). Of course that the alluded helicity has  disturbing effects on the  components of  $\vec A$ inside the coils. Note that the mentioned helicity-effects can be diminished (and practically eliminated)  by using an idea noted in another context in \cite{18}. The respective idea proposes to arrange the spirals of the coil in an even number of layers, the   spirals from adjacent layers having equal helicity but of opposite sense.


\begin{thebibliography}{99}
\bibitem{1}Y. Aharonov and D. Bohm, Significance of electromagnetic potentials in the quantum theory,         \textit{Phys.Rev.}\textbf{115} (1959) 485-491. 
\bibitem{2}Y. Aharonov, D. Bohm,  Further Considerations on Electromagnetic Potentials in the 
Quantum Theory, \textit{Phys.Rev.} \textbf{123} (1961)1511-1524. 
\bibitem{3}.  S.Olariu, I. I.Popescu,  The quantum effects of electromagnetic fluxes, 
\textit{Rev. Mod. Phys.} \textbf{57} (1985) 339-436.
\bibitem{4}M. Peshkin, A. Tonomura, The Aharonov-Bohm Effect, \textit{Lecture Notes in Physics}   
(Springer) \textbf{340} (1989) 1-152.
\bibitem{5} M. Dennis, S. Popescu, L. Vaidman, Quantum Phases: 50 years of the Aharonov-Bohm effect and 25 years of the Berry phase, \textit{J. Phys. A: Math. Theor.} \textbf{43} (2010) 350301
\bibitem{6} A. Ershkovich, Electromagnetic potentials and Aharonov-Bohm effect, arXiv:1209.1078v2, last revised 10 Apr 2013 
\bibitem{7}V. A. Leus,  R.T. Smith, S. Maher, The Physical Entity of Vector Potential in Electromagnetism,  \textit{Appl. Phys.Research},\textbf{ 5 }( 2013) 56 - 68.
\bibitem{8}  B. J. Hiley, The Early History of the Aharonov-Bohm Effect, arXiv:1304.4736v1.
\bibitem{9} Images for G.P. Thomson experiment:\\
\verb'<https://www.google.ro/search?q=g+p+thomson+experiment>'
\verb'<&tbm=isch&tbo=u&source=univ&sa=X&ei=SHnnUtj2CsiAyAPk1oDwAQ>'
\verb'<&ved=0CDAQsAQ&biw=915&bih=737>' , accesed  1/28/2014.
\bibitem{10}T. A. Arias, G.P. Thomson Experiment \\
\verb'<http://muchomas.lassp.cornell.edu/>'
\verb'<8.04/1997/quiz1/node4.html>' , accesed  1/28/2014.
\bibitem{11} M. Born ,  E. Wolf, \textit{ Principle of Optics, Electromagnetic theory of propagation, interference and diffraction of light}, Seventh (Expanded) Edition (Cambridge University Press, Reprinted 2003)
\bibitem{12} S.M. Blinder, Aharonov- Bohm Effect, From the Wolfram Demonstrations Project
http://www.youtube.com/watch?v=OgDPK5MLVnE,  
accesed 2/28/2014
\bibitem{13} J.D. Jackson ,\textit{ Classical Electrodynamics} (John Wiley, N.Y. 1962)
\bibitem{14} L. Landau, E. Lifchitz, \textit{Theorie des Champs} (Ed. Mir, Moscou 1970) 
\bibitem{15} S. Dumitru,  M, Dumitru,  Are there observable effects of the vector potential? : a suggestion for probative experiments of a new type (different from the proposed Aharonov-Bohm one)  CERN Central Library PRE 24618 , Barcode 38490000001, Jan 1981. - 20 p.
\bibitem{16} R.P. Feynman, R.B. Leighton, M. Sands, \textit{The Feynman Lectures on Physics, vol. II} (Addison-Wesley, Reading Mass. 1964).
\bibitem{17}I.S. Gradshteyn, I.M. Ryzhik, \textit{Table of Integrals, Series, 
and Products, Seventh Edition} ( Elsevier,2007).
\bibitem{18}O. Costa de Beauregard,  J. M. Vigoureux, Flux quantization in "autistic" magnets, \textit{Phys. Rev. D} \textbf{9}(1974) 16261632.
\end{thebibliography}
\end{document}